\documentclass[journal]{IEEEtran}
\pdfoutput=1 
\usepackage[pdftex]{graphicx}
\usepackage{amsmath}
\usepackage{url}
\usepackage{booktabs}
\usepackage[caption=false,font=footnotesize]{subfig}
\usepackage{mathtools}
\usepackage{blkarray, bigstrut}
\usepackage[scientific-notation=true]{siunitx}

\newcommand{\AND} {\texttt{AND} }
\newcommand{\IF}  {\texttt{IF} }
\newcommand{\THEN} {\texttt{THEN} }
\begin{document}
%
\title{Rough Set Microbiome Characterisation}

%

\author{Benjamin~Wingfield,
        Sonya~Coleman~\IEEEmembership{Member,~IEEE},        
        T.~M.~McGinnity~\IEEEmembership{Senior Member,~IEEE},\\        
        and Anthony~J.~Bjourson

\thanks{Manuscript received XXX-XXX.}
\thanks{B. Wingfield, S. Coleman, and T. M. McGinnity are with the Intelligent Systems Research Centre, Ulster University at 
L'Derry, U.K.}
\thanks{Anthony~J.~Bjourson is with the Northern Ireland Centre for Stratified Medicine, Ulster University at L'Derry, U.K.}%
}%

\markboth{Journal of \LaTeX\ Class Files,~Vol.~14, No.~8, August~2015}%
{Shell \MakeLowercase{\textit{et al.}}: Bare Demo of IEEEtran.cls for IEEE Journals}
%


\maketitle

\begin{abstract} Microbiota profiles measure the structure of microbial
communities in a defined environment (known as microbiomes). In the past decade,
microbiome research has focused on health applications as a result of which the gut microbiome has
been implicated in the development of a broad range of diseases such as obesity,
inflammatory bowel disease, and major depressive disorder. A key goal of many microbiome experiments
is to characterise or describe the microbial community. High-throughput
sequencing is used to generate microbiota profiles, but data gathered via this
method are extremely challenging to analyse, as the data violate multiple strong
assumptions of standard models. Rough Set Theory (RST) has weak assumptions that are less likely to be violated, and offers a range of attractive tools for extracting knowledge from complex data.
In this paper we present the first application of RST for characterising
microbiomes. We begin with a demonstrative benchmark microbiota profile and
extend the approach to gut microbiomes gathered from depressed subjects
to enable knowledge discovery. We find that RST is capable of excellent
characterisation of the gut microbiomes in depressed subjects and
identifying previously undescribed alterations to the microbiome-gut-brain axis.
An important aspect of the application of
RST is that it provides a possible solution to an open research question regarding the search for an optimal normalisation approach for microbiome census data, as one
does not currently exist.
\end{abstract}

%
\IEEEpeerreviewmaketitle

\section{Introduction} \label{sec:introduction}

High-throughput DNA sequencing has enabled culture-independent profiling of
complex microbial communities. A microbiota is defined as the assemblage of
micro-organisms present in a defined environment \cite{marchesi2015vocabulary}.
Over the past decade interest about the role microbiota play in both health and
disease, across the human body, has rapidly grown. Microbes directly and
indirectly interact with human physiology through a variety of mechanisms,
including protecting against pathogenic infections, contributing to normal
metabolic functions, and by training the immune system \cite{shreiner2015gut}.
Alterations to the gut microbiota have been implicated in the pathogenesis of a
number of diseases, including Inflammatory Bowel Disease, diabetes, obesity, and
depression \cite{gevers2014treatment, hartstra2015insights, zheng2016gut}.
Typically a microbiota profile is created by collecting an environmental sample
(e.g.\ stool to study the human gut microbiota), extracting the DNA present, and
sequencing the 16S ribosomal RNA (16S rRNA) marker gene. Put simply, the 16S
rRNA marker gene is similar to a barcode: different bacterial species have
different marker gene sequences. Marker gene sequences can be matched to
reference databases to discover the taxonomy of a given sequence (e.g. \texttt{ACTG...} $\mapsto$ \textit{E. coli}). A
high-throughput DNA sequencer will create millions of discrete DNA sequence
reads, often between 200 -- 400 nucleotides long. Microbiota profiles (also
known as microbiome census data) are extremely challenging to analyse: they are
highly dimensional, sparse, noisy, and compositional. These properties violate
many assumptions of standard models that have been widely applied to analyse
microbiota profiles \cite{mcmurdie2014waste} (described further in
Section~\ref{sec:background}). A broad range of complex normalisation algorithms
(data transformations) has been developed and applied to resolve the
problems inherent to such data \cite{weiss2017normalization}. However,
identifying which normalisation algorithm is optimal remains an open question in
the microbiome research community. In contrast, data-driven computational
intelligence paradigms have minimal or weak prior assumptions and offer powerful
tools for extracting knowledge from problematic data such as microbiota profiles
\cite{lahat2015multimodal}. As such, their application to microbiome census data may offer the potential for more robust analysis. 

Among computational intelligence/AI approaches, Rough Set Theory (RST) is a topic of great interest amongst the research
community and has been applied to a variety of domains for the purpose of data
analysis \cite{pawlak1998rough}, including many areas of computational biology
\cite{petit2014rough}. Many microbiome
experiments aim to identify correlations between the characterised microbial
community and disease. Only a small part of this process is concerned with
evaluating predictive power: the process of determining elements of a microbial
community that have predictive power is described as biological marker
(biomarker) analysis by molecular biologists. RST can elegantly address both
characterisation and prediction. Firstly, by identifying a minimal knowledge
representation (a reduct), redundant or irrelevant bacterial species can be
discarded, simplifying analysis. Additionally, transparent rules can be induced
to describe the minimal knowledge representation, enabling knowledge discovery.
If additional data are available the rules can be used to evaluate the
predictive power of putative biomarkers. A combination of both approaches allows
domain experts to interpret a model and gain an understanding of the underlying
biological processes involved. Additionally, RST does not require parameters to
be set, which eliminates a source of potential human bias.

This paper presents an approach to characterise microbial communities using RST
to simultaneously transform data into knowledge and to resolve an open research
question regarding normalising microbiome census data (described further in
Section~\ref{sec:background}). This approach is first demonstrated on a small
benchmark dataset to characterise the microbial communities present across
different human body sites. This serves as a simple demonstration of RST, because
it is well known that microbial communities significantly differ across the
human body. The application of RST is then expanded to cover a microbiome
experiment that investigates the link between microbiomes and depression (see
Section~\ref{sec:background}) to enable knowledge discovery. It is important to
note that this paper focuses on characterisation and does not address prediction
from the induced rules (i.e.\ the rules are descriptive and reveal underlying
patterns in the data). The rationale for this is that characterisation enables
the transformation of data into knowledge. From this knowledge insights about
microbial communities can be gained and future experiments planned. The quality
of characterisation can be measured by a variety of measures, discussed in
Section~\ref{sec:background}. The application of RST enables microbial
ecologists to understand better \textit{what} is happening in a microbial community ---
by removing the consideration of superfluous bacterial species and the requirement for destructive
data transformations --- and \textit{why} bacterial species are associated with
phenotypes, via the analysis of transparent induced rules.

The remainder of this paper is structured as follows:
Section~\ref{sec:background} describes some problematic properties of microbiome
census data that make analysis challenging, the links between the microbiome and depression, and core RST
concepts applied throughout this work. Section~\ref{sec:methods} introduces the
rough set microbiota model and describes the datasets used to benchmark the
model. An evaluation of the rough microbiota model that demonstrates the potential of applying RST to
microbiota profiles is provided in Section~\ref{sec:results}. Finally, conclusions and an outline for future work are presented in Section~\ref{sec:conclusion}.

\section{Background} \label{sec:background}

Machine learning and computational intelligence approaches are often applied to
microbiome census data for the purpose of predicting a categorical or numeric
variable from a set of input data (e.g.\ predicting disease). However,
classification and regression are only a subset of data mining and knowledge
discovery tasks. Popular tasks for data mining and knowledge discovery include
\cite{larose2014discovering}:

\begin{itemize}
  \item Describing patterns and trends in data;
  \item Approximating a categorical target variable from a larger data set (classification); 
  \item Approximating a numeric target variable from a larger data set (regression);
  \item Predicting future events (e.g.\ the share price of a company in 3
    months);
  \item Clustering observations into similar groups;
  \item Identifying association rules (finding features that co-occur).
\end{itemize}

\noindent Describing patterns and trends in data is the most common aim of
microbiome experiments. Many microbiome experiments aim to identify correlations
between the characterised microbial community and disease. The process of
determining elements of a microbial community that have predictive power is
described as biological marker (biomarker) analysis by molecular biologists. When attempting to describe patterns and trends in data, models that offer transparency provide significant benefits \cite{larose2014discovering}; RST provides a suite of tools that allows the transparent description of data, which could help to fulfil an important objective for molecular ecologists.

\subsection{Why RST?: Problematic properties of microbiota profiles}

Microbiome census data produced by high-throughput sequencing are extremely
challenging to analyse: they are highly dimensional
\cite{statnikov2013comprehensive}, noisy \cite{callahan2016dada2}, variably
sparse across different environments \cite{paulson2013differential},
compositional \cite{gloor2016compositional}, and have an uneven mean-to-variance
relationship \cite{mcmurdie2014waste}. Most of these properties will violate the
assumptions of standard analysis models, such as normality or homoscedasticity. For example, investigating if certain bacterial species are more or less abundant in certain environments is difficult because varying sparsity across
different environments can violate probability distribution assumptions. As microbiome census data are not normally distributed and are heteroscedastic it is not appropriate to model differential abundance with popular approaches such as $t$-tests. Microbial ecologists interested in examining bacterial co-occurrence relationships will have problems computing correlation coefficients  \cite{friedman2012inferring}. A thorough review discussing these problems is available
\cite{weiss2017normalization}. After initial quality control and clustering
preprocessing steps \cite{kozich2013development} (or alternatively denoising
\cite{callahan2016bioconductor}) microbial community sequencing data are
typically organised into large matrices where rows represent samples and columns
represent counts of clustered sequence reads that constitute different types of
bacteria. The number of discrete sequence reads per sample (the sum of each row)
can differ by orders of magnitude. This uneven sampling effort does not reflect
true biological variation and is an artefact of the sequencing process. The
uneven sampling effort will bias the estimates of bacterial abundance and should
be normalised to allow fair comparison between samples. Normalisation procedures
can also mitigate the other types of bias present in microbial community
sequencing data described above. However, recommended normalisation procedures
that aim to mitigate such complex problems are often difficult for
microbiologists to incorporate (e.g.\ applying a variance stabilising
transformation based on Gamma-Poisson mixture models \cite{love2014moderated})
and can destroy the semantics of the original data. A widely used normalisation
strategy is to convert counts into relative abundances per sample (simple
proportions). However, as relative abundances are constrained by an artificial
limit (1) they represent compositional data. Compositional data have an
arbitrary or non-informative sum \cite{aitchison2005compositional}.
Additionally, relative abundance data can be skewed by the presence of highly
abundant species, and the transformation does not resolve important problems
with the data such as heteroscedasticity \cite{mcmurdie2014waste}. However, it
is rare for microbial communities in the human body to be dominated by a few
species, and relative abundance data are easy for microbial ecologists to
incorporate and analyse. Crucially, the problematic properties of relative
abundance data (heteroscedasticity and compositionality) do not violate the
assumptions of RST, which is our motivation for using this type of
normalisation.

The only assumption required in RST is that each object has an associated set of
attributes used to describe the object, and that the data are a true and
accurate reflection of reality \cite{jensen2008computational}. Thus, the
application of RST resolves the problematic aspects of microbiome census data
described above. Extensive steps were taken to denoise the microbiome census
data, described further in Section III, to ensure that the accuracy assumption
is not violated. Additionally, RST makes redundant the requirement for more complex
normalisation algorithms; the semantics of easily intuited relative abundance
microbiome census data are maintained --- aiding interpretation by domain
experts and providing a possible solution to an open question in the microbiome
research community regarding choice of an optimal normalisation algorithm (which
can differ depending on data and analysis task \cite{weiss2017normalization}).

As far as can be ascertained, there have been no previous attempts to perform
data analysis on microbiome census data using RST described in the literature.
However, aspects of RST have been implemented for bioinformatics applications in
the wider field of metagenomics. The metagenome is defined as the collection of
genomes and genes from the members of a microbiota
\cite{marchesi2015vocabulary}. RST has been applied to remove superfluous
$K$-mers and to improve DNA fragment classification compared with standard
bioinformatics tools \cite{jian2015reduction}. A rough reduction method based on
Particle Swarm Optimisation has also been applied to the same problem
\cite{jian2016rough}. RST has also been used to predict the presence of operons
in metagenomic data. A decision tree classifier based on the Variable Precision
Rough Set Model (VPRSM) was applied to genomic data from \textit{Escherichia
  coli} to identify if a gene belonged to an operon
\cite{zaidi2016computational}. The VPRSM had an accuracy of 89.4\% using five
features: maximum distance, minimum distance, direction, cluster of orthologous
groups, and gene order conservation. The use of a decision tree meant that the
decisions of the classifier were easy to interpret and could be validated by
domain experts.

\subsection{The microbiome and depression}

Depression is a mental disorder that causes a persistent low mood, low self
esteem, and chronic anhedonia. Depression is currently the leading cause of 
global disease and affects over 300 million people worldwide
\cite{world2017depression}. A growing body of evidence supports the hypothesis that the gut microbiota, the complex community of microorganisms that inhabit the human
gastrointestinal tract, play a key role in the aetiology of
depression via regulation of the central nervous system in a complex network
known as the microbiome-gut-brain axis; an in depth explanation of this
phenomena is outside the scope of this paper, but comprehensive reviews are
available on the subject \cite{cryan2012mind, foster2013gut}. Despite work in
animal models that links the gut microbiota and depression
\cite{park2013altered}, limited work has been done using a human cohort: faecal
samples isolated from Norweigan \cite{naseribafrouei2014correlation} and Chinese
\cite{jiang2015altered, zheng2016gut} cohorts have identified some alterations
in a depressed cohort. Our rationale for applying RST to a publicly available
depression microbiome dataset, described further in Section~\ref{sec:methods},
is to enable the extraction of new knowledge from the data as previous work has
relied on standard analysis techniques.

\subsection{Core RST concepts}

A microbiota profile can be represented by a $M \times N$ decision table. The
rows of a decision table correspond to the universe of discourse, $X$
\cite{jensen2008computational}:

\begin{equation}
X = \{x_1, x_2, \ldots, x_N\}
\end{equation}

\noindent The columns of a decision table correspond to the set of features $A$
(the set of microbes) \cite{jensen2008computational}:

\begin{equation}
A = \{a_1, a_2, \ldots, a_M\}
\end{equation}

\begin{figure}[t]
  \centering
  \includegraphics[width=0.8\columnwidth]{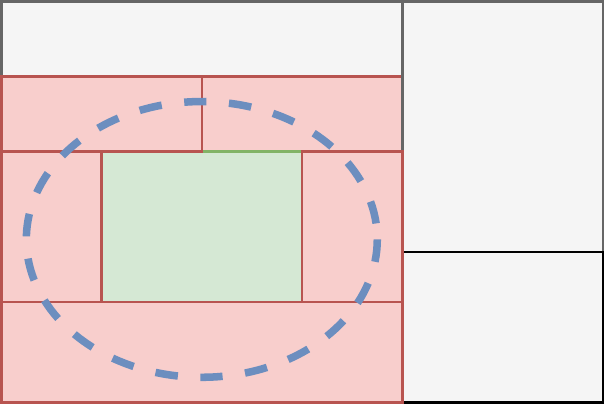}
  \caption{Rough set example. The universe of discourse is partitioned into 9
    indiscernibility classes by a set of attributes. The blue line represents
    the set being approximated (e.g.\ sick subjects). The green section is the
    lower approximation, and the red sections are the upper approximations of
    the rough set. In the complement of the upper approximation (grey) it is
    certain that no objects in the rough set will be present (e.g.\ a healthy
    subject could be in the grey section)}
  \label{fig:rs}
\end{figure}

\noindent Decision table $DT$ consists of a subset of condition attributes
(input features, different microbial species) and decision attributes (class
labels e.g.\ disease or healthy; $DT = C \cup D$). Each attribute has an
associated value set, which represents the abundance of the microbial species
\cite{petit2014rough}:

\begin{equation}
  V_a = \{v^a_1, v^a_2, \ldots, v^a_p\}
\end{equation}

\noindent where $a \in A$. The value set must be discrete (continuous variables
must be discretised). Although microbiome census data are
discrete counts of sequences, they are typically converted into continuous
variables by a normalisation process to mitigate uneven library size bias. A thorough discussion of bias in microbiome census data and normalisation approaches to counteract this is available \cite{weiss2017normalization}.
Therefore relative abundance microbiome census data, which is used as input data
throughout this chapter, must first be discretised. The maximal discernibility
heuristic was used to discretise the microbiome census data throughout this
paper \cite{bazan2000rough}. Any condition or decision attribute subset $P \subseteq C\
\text{or}\ D$ can induce a partition in $X$ \cite{petit2014rough}:

\begin{equation}
X \xrightarrow{P} X(P) = \{ X_1^P, \ldots, X_q^P\}
\end{equation}

\noindent where $X_l^P$ is the partition of $X$ induced by $P$. The subsets
\cite{petit2014rough}:

\begin{equation}
X = X_a^P \cup \ldots \cup X_Q^P
\end{equation}

\noindent correspond to the set of equivalence classes, called indiscernibility
classes in RST. \noindent Discernibility is the core concept of RST: if $(x, y)
\in \text{IND}(P)$ (where $\text{IND}(P)$ is the indiscernibility relation
induced by attribute subset $P$) then $x$ and $y$ are indiscernible by
attributes from $P$. For example, if two bacterial species have the same
abundance in both healthy and sick subjects, then using only the abundance of
the bacterial species it is impossible to discern between the two subjects. In
RST, a set is approximated by two sets known as the lower and upper
approximations \cite{jensen2008computational}:

\begin{align}
\underline{P}S = \{x:[x]_P \subseteq S \} \\
\bar{P}S = \{x: [x]_P \cap S \neq \emptyset \}
\end{align}

\noindent where $S \subseteq X$ and $[x]_P$ are the equivalence classes of the
$P$-indiscernibility relation. The tuple $\langle\underline{P}S,
\bar{P}S\rangle$ is known as a rough set. $P$ and $Q$ are sets of attributes
inducing equivalence relations over $U$. The region between the upper and lower
approximation sets is called the boundary region. The boundary region represents
the set of objects that can possibly be predicted to be from a specific decision
class (non-deterministic; see Figure~\ref{fig:rs})
\cite{jensen2008computational}:

\begin{align}
  \text{BND}_P(Q) = \bigcup_{X \in U/Q} \bar{P}S - \bigcup_{X \in U/Q} \underline{P}S
\end{align}

\noindent The positive region, in which objects can be predicted to belong to a
decision class with certainty, is given by:

\begin{align}
\text{POS}_P(Q) = \bigcup_{X \in U/Q} \underline{P}Y \\
\end{align}

\noindent The negative region represents the set of objects that cannot be
predicted to a decision class:

\begin{align}
\text{NEG}_P(Q) = X - \bigcup_{Y \in X / Q} \bar{P}Y 
\end{align}

\noindent Attributes that cannot be removed without changing the partitioning of
objects amongst the indiscernibility relations are indispensable. A minimal set
of indispensable condition attributes is known as a reduct.

\section{Rough microbiome analysis} \label{sec:methods}

\begin{figure*}[!t]
  \centering
  \includegraphics[width=1\linewidth]{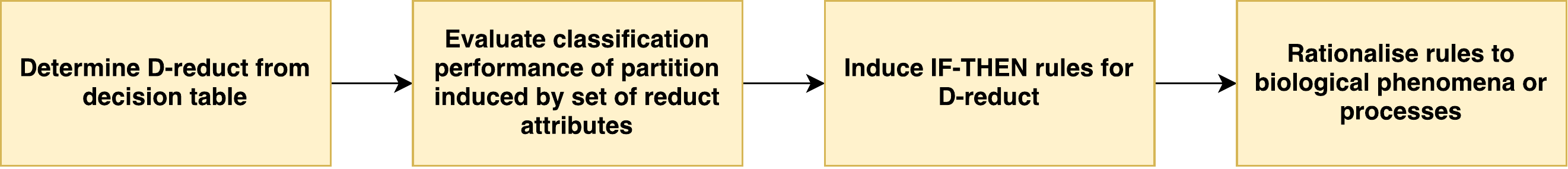}
  \caption{Overview of rough microbiota profile analysis}
  \label{fig:overview}
\end{figure*}

This work uses two datasets:

\begin{enumerate}
  \item a publicly available human body site dataset \cite{caporaso2011global},
    which contains environmental samples gathered from across the human body
    (tongue, skin, or faeces);
  \item a publicly available gut microbiome depression dataset 
  \cite{jiang2015altered}, which contains faecal samples gathered from depressed and control subjects.
\end{enumerate}

The human body site dataset serves to demonstrate the RST approach before it is
applied to more complex data for knowledge discovery. The human body site
dataset contains 3 samples for each body site: human skin, human tongue, and
human faeces (9 samples total). This dataset forms part of the larger ``Global
Patterns'' dataset --- in microbiome research the Global Patterns dataset is
widely used to benchmark new algorithms or tools \cite{mcmurdie2014waste,
  weiss2017normalization}, which is the rationale for applying RST to these
data. The human body site dataset was analysed using the \texttt{RoughSets}
package \cite{riza2014implementing} implemented in \texttt{R}.

The public gut dataset consists of a cohort containing 59 faecal samples (30 control, 29 depressed). Briefly, the bacterial DNA present in the samples was extracted and sequenced with a 16S marker gene survey. This process generated millions of DNA sequences, 200--400 nucleotides in length. Before this raw sequence data can be input to  the rough set microbiome characterisation process, it must be first processed with bioinformatics algorithms to generate an accurate survey of bacterial species.  The gut data were denoised according to standard operating protocols using an \texttt{R} bioinformatics pipeline \cite{callahan2016bioconductor} to ensure that the truth assumption of RST was met. The human body site dataset was input to the rough set microbiome characterisation process in its preprocessed form, which is often used to to simplify analysis. The data were discretised with
the maximum discernibility discretisation algorithm implemented in the
\texttt{Java} \texttt{rseslib} library \cite{bazan2000rses}. Due to the scale of the depression datasets (both contained 3000 -- 4000 features) the rough set microbiome characterisation was implemented with  \texttt{rseslib}, as \texttt{R} (the language in which the \texttt{RoughSets} package was implemented)
suffers from poor computational performance compared with other languages such as
\texttt{Java} \cite{wickham2014advanced}. To aid analysis, discrete data in the
depression datasets were labelled low and high if the bacterial species
abundance had two cuts, and low, medium, and high if the bacterial species
abundance had three cuts. We evaluated the ability of RST to model microbiota
profiles by testing classification performance on the following tasks:

\begin{enumerate}
\item Classify microbial communities from different areas of the human body
  (tongue, skin, or faeces);
\item Classify depression status from the gut microbiome.
\end{enumerate}

We began by generating a single reduct for the first decision table (the
demonstrative dataset; see Figure~\ref{fig:overview}). The rationale for
generating a single reduct is that the first decision table serves as a
demonstration of RST applied to a simple problem, and a single reduct can be
used to provide a simple description of microbial communities across the human
body. For the depression datasets, all local reducts were computed for each
decision table using the \texttt{rseslib} \texttt{Java} library. To determine
the classification performance of the partition in $X$ induced by the set of
reduct attributes $A_k$ two measures were used \cite{petit2014rough}:

\begin{align}
\label{eq:acc}
  \text{Accuracy} = \frac{\sum_{L=1}^Q \text{Card}(\underline{A}_k X^{A_k}_L)}{\sum_{L=1}^Q \text{Card}(\bar{A}_k X^{A_k}_L)} \\
  \text{Quality} = \frac{\sum_{L=1}^Q \text{Card}(\underline{A}_k X^{A_k}_L)}{\text{Card}(X)}
\label{eq:qual}
\end{align}

\noindent where $\text{Card}$ is cardinality, which represents the number of
elements in a set, and $L$ is the total number of upper ($\bar{A}_kX_L^{A_k}$)
and lower-approximation ($\underline{A}_kX_L^{A_k}$) set tuples. Accuracy
represents the ratio of the size of all lower-approximation sets to the size of
all upper-approximation sets ($0 \leq \text{Accuracy}[X(A_k)]\leq 1$). If the
family of lower approximation sets is an empty set (i.e. no objects can be said
to be certainly predicted) then accuracy is zero. Quality represents the ratio
of all objects in the family of lower approximation sets to the total number of
objects in the universe of discourse ($0 \leq \text{Quality}[X(A_k)]\leq 1$). It
is important to note that classification accuracy and quality are not tested on
independent validation data due to insufficient data: these metrics are only
capable of explaining how well a rough set is describing a microbiota profile.
\texttt{IF-THEN} decision rules were generated from the indiscernibility classes
defined by the reduct attributes using the \texttt{rseslib} library. The
descriptive strength of the rules was evaluated by measuring the support each
rule has. Once species of interest were identified by this procedure, a literature review was conducted to identify associations between the generated rules and biological phenomena. 

\section{Results} \label{sec:results}

\subsection{Human body site data} 
 
\begin{table}[t]
\centering
\caption{Classification metrics}
\label{tab:metrics}
\begin{tabular}{@{}lll@{}}
\toprule
Classification task & Accuracy & Quality \\ \midrule
Human body site     & 1        & 1       \\
Gut microbiome      & 1        & 1       \\ \bottomrule
\end{tabular}
\end{table}

A decision table was created from the human body site data which contained 3878 conditional attributes and 9 samples (3 skin samples, 3 faecal samples, and 3 tongue samples). Each conditional attribute defines the abundance of a bacterial Operational Taxonomic Unit, which approximates a bacterial taxa (group e.g. species). A single reduct was generated for this first characterisation task as it serves as a demonstration before the RST approach is expanded to the depression data set for knowledge discovery. A single feature was present in the reduct: the bacterial species
\textit{Propionibacterium acnes}. The characterisation ability of the reduct
rough set was tested using the accuracy and quality measures described in
Equations \ref{eq:acc} and \ref{eq:qual} (see Table~\ref{tab:metrics}). The
lower approximation set contained all of the samples for each sample type so the
accuracy and quality of classification was 1. This demonstrates that RST is capable of excellently discerning between samples collected from different sites across the human body. The next step of characterisation is to describe the alterations identified by RST using \texttt{IF-THEN} rules and linguistic variables.

Rules were generated from the reduct. It is important
to note that the generated rules are descriptive and the predictive power of
them is not assessed: although prediction can be valuable, it forms only one
aspect of a microbiome experiment. Patterns and trends in data can be described
by generating and analysing a set of \texttt{IF-THEN} rules. However, the
quality of characterisation can be measured by the strength of the generated
rules, which is defined as the number of instances in the dataset that are
concordant with each rule. The human body site characterisation task
generated three rules (with 100\% strength) for three classes regarding the
bacterial species \textit{P. acnes}:

\begin{align} 
\label{eq:third-start} \tag{Rule 1}
\text{\texttt{IF} \textit{P. acnes }}  0 \text{\texttt{ THEN} Faeces} \\ 
\tag{Rule 2}
\text{\texttt{IF} \textit{P. acnes }}  [0,\num{4.91e-05}] \text{\texttt{ THEN} Tongue} \\
\label{eq:third-end}
\tag{Rule 3}
\text{\texttt{IF} \textit{P. acnes }}  [\num{4.91e-05},1] \text{\texttt{ THEN} Skin}
\end{align}

\noindent The generated rules are supported by compelling biological evidence.
Relating the output of the RST characterisation process to biological phemonena
is simple because the semantics of the original data were not destroyed by
complex normalisation approaches. Typically \textit{P. acnes} is a commensal
member of the skin microbiome, but it can act as a pro-inflammatory
opportunistic pathogen, causing acne \cite{perry2011propionibacterium}. Its
pattern of abundance matches descriptions in the literature: most prevalent on
skin, but capable of colonising other areas of the body including the tongue and
large intestine \cite{perry2011propionibacterium}. The absence of \textit{P.
acnes} in stool samples could be related to the sensitivity of the sequencing
process or the low sample size of the cohort (\textit{P. acnes} is not a major
member of the gut microbiome, the most complex of all human microbiomes).
Alternatively, as faeces are not a perfect proxy for the large intestine
\textit{P. acnes} may be present in the large intestine but be undetectable in
stool samples. The RST characterisation of human body sites has therefore identified a
biologically plausible process that represents a key change across habitats. We
will now apply this approach to the more complex depression dataset to enable
knowledge discovery.

\subsection{Gut data} 

A decision table created from the gut dataset using the approach described in Section~\ref{sec:methods}. The gut decision table had 2900 conditional attributes, and 59 samples
(30 control, 29 depressed). Each conditional attribute defines the abundance of a denoised amplicon sequence variant which approximates the true DNA sequence present in the samples. The denoising paradigm offers a range of benefits --- including sampling accuracy --- compared with the operational taxonomic unit approach, which is our motivation in using the approach. A thorough explanation of the benefits the denoising paradigm offers is outside the scope of this paper;  reviews are available \cite{callahan2016dada2}; the denoising approach did not exist when the human body site data were first created. All local reducts were computed for the gut decision table. The reducts contained 12 features that covered the bacterial genera \textit{Bacteroides}, \textit{Prevotella}, \textit{Anaerostipes},
\textit{Phascolarctobacterium} and \textit{Odoribacter}. One of the features
could not be mapped to a specific genus, and represented the bacterial family
\textit{Ruminococcaceae}. The lower approximation set contained all of the samples for both classes
(depression and control), so the accuracy and quality of characterisation was 1
(creating a crisp set), which demonstrates that the rough set microbiome characterisation is capable of excellently discerning between samples collected from depressed and control subjects. 

The next step of characterisation is to describe
the alterations identified by RST using \texttt{IF-THEN} rules and linguistic
variables. More complex rules were generated to characterise the gut microbiome characterisation task (see
Table~\ref{tab:gut-diagram}). The abundance of bacterial taxa was defined as being low or high to aid
comprehension. In the gut microbiome three rules were induced to characterise
control samples, and four rules to characterise depressed samples. Control
samples are characterised by low abundance of the bacterial genera, whilst
depressed samples are characterised by a mixture of high and low abundant
bacterial genera. There are significant underpinning biological justifications for the four
rules that characterise the depressed gut microbiome.
\textit{Phascolarctobacterium} is a bacterial genus that is abundant in the
human gut and produces short chain fatty acids, which are associated with
modifying host metabolism and mood \cite{cryan2012mind}. Additionally,
\textit{Phascolarctobacterium} has been previously positively correlated with
positive mood in healthy adults \cite{li2016gut}. The second and third rules for
depression contain the multiple bacteria in the \textit{Bacteroides} genus;
\textit{Bacteroides} are a major mutualistic member of the normal human
intestinal microbiome, the described abundance patterns indicate a type of gut
dysbiosis has occurred, which has been frequently associated with various diseases \cite{rogers2016gut}.

It is useful to compare our results for the gut microbiome with the original
analysis that used a traditional (i.e. non-RST) methodology
\cite{jiang2015altered}. The low levels of \textit{Ruminococcaceae} in rules 2
and 3 are concordant with the traditional analysis. The low abundance of
\textit{Alistipes} in rule 4 is not consistent with the traditional analysis.
However, the low abundance is combined with a high abundance of
\textit{Odoribacter}, which was also not mentioned in the original analysis.
\textit{Odoribacter} are typically opportunistic pathogens, which can activate
inflammatory pathways associated with the microbiome-gut-brain axis
\cite{hardham2008transfer}. 

\begin{table*}[t]
\centering
\caption{Rules that characterise the gut microbiome. $\Uparrow$ indicates high
abundance, $\Leftrightarrow$ indicates medium abundance, and $\Downarrow$
indicates low abundance}
\label{tab:gut-diagram}
\begin{tabular}{p{1cm}p{1cm}p{8cm}p{1cm}p{2cm}}
\toprule
Rule &     & Antecedent                                                                                                                                                                                                                                                                                         &       & Consequent  \\ \midrule
1    & \IF & \textit{Bacteroides} (3) $\Downarrow$ \AND \textit{Bacteroides} (6) $\Downarrow$ \AND \textit{Prevotella} $\Downarrow$ \AND \textit{Anaerostipes} $\Downarrow$ \AND \textit{Ruminococcaceae} $\Downarrow$                                                                                          & \THEN & Control     \\
2    & \IF & \textit{Bacteroides} (3) $\Downarrow$ \AND  \textit{Bacteroides} (4) $\Downarrow$ \AND \textit{Bacteroides} (6) $\Downarrow$ \AND  \textit{Anaerostipes} $\Downarrow$ \AND  \textit{Ruminococcaceae} $\Downarrow$                                                                                  & \THEN & Control     \\
3    & \IF & \textit{Bacteroides} $\Downarrow$ \AND  \textit{Bacteroides} (1) $\Downarrow$ \AND  \textit{Bacteroides} (4) $\Downarrow$ \AND  \textit{Bacteroides} (6) $\Downarrow$ \AND  \textit{Ruminococcaceae} $\Downarrow$ \AND  \textit{Odoribacter} $\Downarrow$ \AND  \textit{Anaerostipes} $\Downarrow$ & \THEN & Control     \\ \cmidrule{2-5}
1    & \IF & \textit{Bacteroides} (1) $\Downarrow$ \AND  \textit{Bacteroides} (6) $\Uparrow$ \AND  \textit{Phascolarctobacterium} $\Downarrow$ \AND  \textit{Ruminococcaceae} $\Downarrow$                                                                                                                      & \THEN & Depressed   \\
2    & \IF & \textit{Bacteroides} (3) $\Downarrow$ \AND  \textit{Ruminococcaceae} (6) $\Downarrow$ \AND  \textit{Bacteroides} (6) $\Uparrow$                                                                                                                                                                    & \THEN & Depressed   \\
3    & \IF & \textit{Bacteroides} (1) $\Downarrow$ \AND  \textit{Bacteroides} (4) $\Downarrow$ \AND  \textit{Bacteroides} (6) $\Uparrow$ \AND  \textit{Ruminococcaceae} $\Downarrow$                                                                                                                            & \THEN & Depressed   \\
4    & \IF & \textit{Alistipes} $\Downarrow$ \AND  \textit{Odoribacter} $\Uparrow$                                                                                                                                                                                                                              & \THEN & Depressed   \\ \cmidrule{2-5}
1    & \IF & \textit{Bacteroides} (3) $\Leftrightarrow$ \AND  \textit{Bacteroides} (6) $\Downarrow$ \AND  \textit{Odoribacter} $\Downarrow$ \AND  \textit{Oscillospira} $\Downarrow$ \AND  \textit{Anaerostipes} $\Downarrow$                                                                                   & \THEN & Remission   \\
2    & \IF & \textit{Bacteroides} (3) $\Leftrightarrow$ \AND  \textit{Bacteroides} (4) $\Downarrow$ \AND  \textit{Phascolarctobacterium} $\Downarrow$ \AND  \textit{Oscillospira} $\Downarrow$                                                                                                                  & \THEN & Remission   \\
3    & \IF & \textit{Bacteroides} $\Downarrow$ \AND  \textit{Bacteroides} (3) $\Leftrightarrow$ \AND  \textit{Bacteroides} (6) $\Downarrow$ \AND  \textit{Odoribacter} $\Downarrow$                                                                                                                             & \THEN & Remission   \\
4    & \IF & \textit{Bacteroides} $\Downarrow$ \AND  \textit{Bacteroides} (1) $\Uparrow$ \AND  \textit{Odoribacter} $\Downarrow$                                                                                                                                                                                & \THEN & Remission   \\
5    & \IF & \textit{Bacteroides} $\Downarrow$ \AND  \textit{Bacteroides} (1) $\Uparrow$ \AND  \textit{Bacteroides} (3) $\Leftrightarrow$                                                                                                                                                                       & \THEN & Remission   \\
6    & \IF & \textit{Bacteroides} (3) $\Leftrightarrow$ \AND  \textit{Phascolarctobacterium} $\Downarrow$ \AND  \textit{Odoribacter} $\Downarrow$ \AND  \textit{Oscillospira} $\Downarrow$                                                                                                                      & \THEN & Remission   \\
7    & \IF & \textit{Bacteroides} (3) $\Leftrightarrow$ \AND  \textit{Bacteroides} (6) $\Downarrow$ \AND  \textit{Ruminococcaceae} $\Downarrow$ \AND  \textit{Odoribacter} $\Downarrow$                                                                                                                         & \THEN & Remission   \\
8    & \IF & \textit{Bacteroides} (1) $\Uparrow$ \AND  \textit{Bacteroides} (3) $\Downarrow$ \AND  \textit{Odoribacter} $\Uparrow$ \AND  \textit{Oscillospira} $\Leftrightarrow$                                                                                                                                & \THEN & Remission   \\ \bottomrule
\end{tabular}
\end{table*}

\begin{table}[t]
\centering
\caption{Quality of microbiome characterisation.}
\label{tab:support}
\begin{tabular}{@{}lll@{}}
\toprule
\multicolumn{3}{c}{Gut microbiome}     \\ \cmidrule{1-3} 
Decision  & Rule \# & Support \\ \midrule
Control   & 1       & 86.7\%  \\
          & 2       & 83.3\%  \\
          & 3       & 80.0\%  \\ \cmidrule{2-3}
Depressed & 1       & 31.0\%  \\
          & 2       & 27.5\%  \\
          & 3       & 27.5\%  \\
          & 4       & 27.5\%  \\ \bottomrule
\end{tabular}
\end{table}

\section{Discussion} 

The accuracy and quality of characterisation was 1 for all decision tables
(i.e.\ all objects were in the lower approximation for all decision tables).
Therefore the rough sets constructed from each reduct were able to perfectly
discern samples by class (e.g.\ body site location or depression status) from
the microbiota profiles. The ability to describe clear differences between samples in a transparent and simple way is invaluable for biologists, and applying RST to larger microbiome datasets could help to confirm associations between specific microbial species and disease states.

The use of RST makes obsolete the need for complex
normalisation algorithms. Typically some kind of normalisation is required while
generating microbiota profiles \cite{weiss2017normalization} to allow fair
comparison between samples and to mitigate problematic properties of
high-throughput sequencing data, but the choice of optimal normalisation
algorithm remains an open question. This suggests that RST could be a valuable
tool for modelling human microbiomes. Microbiomes across the human body have
been implicated in the pathophysiology and aetiology of a variety of diseases;
the microbiome plays such an important role in disease and health that it has
been dubbed the ``the second genome'' \cite{grice2012human}. Removing the
requirement for more complex normalisation techniques improves the ability of
biological domain experts to comprehend the output of the characterisation, and
also reduces the computational burden of creating microbiota profiles.

In the gut microbiome the support for control rules (83.3\% average) was higher
compared to the support for depression related rules (28.4\%). The lower support for
depressed rules is in line with current theories regarding the
microbiome-gut-brain axis: it is thought the gut in depressed subjects is in a
state of dysbiosis. Dysbiosis describes microbial imbalance, which can vary
significantly across different subjects. Additionally, microbiome composition
can differ significantly across individuals with dysbiosis, while the overall
gene content may be the same (i.e.\ the functions of the bacteria)
\cite{dash2015gut}. 

\section{Conclusion and limitations} \label{sec:conclusion}

In this work we present the first application of RST to characterise
microbiota profiles from a standard benchmark dataset. We then extend the
application to a gut microbiome dataset to enable
knowledge discovery concerning microbiomes in depressed subjects. We find that
RST is capable of excellently characterising the gut  microbiomes in
depressed subjects and identifying previously undescribed alterations to the
gut microbiome in depressed subjects. The minimal prior assumptions of
RST also offer a potential solution to an unresolved question in the microbiome
community regarding identifying which normalisation procedure is optimal for
microbiota profiles. In addition, the application of RST tools such as reducts
and rule induction allows domain experts to understand RST models without
requiring an understanding of the mathematics involved and thus helps to shed
light on the underlying biological processes present.

Rule-based systems suffer from the combinatorial rule explosion problem. As the
number of features being considered increases, the number of rules increases
exponentially \cite{combs1998combinatorial}. This drastically reduces the
performance and transparency of rule-based systems. The rough set theory
applications to the microbiome census data in this paper have generated small
reducts, with less than a dozen features, which avoids this problem. However, it
is important to note some applications of the rough set characterisation approach may
result in a rule explosion if many bacterial species are relevant to the
characterisation. Rule optimisation would be an important method of tackling
this problem while maintaining the transparency of the system.
 
The biggest limitation to the proposed approach is the
small sample size of the analysed data. Microbial ecologists use a range of measures to ensure that
sampling has been sufficient to enable the accurate characterisation of an
environment (e.g.\ taxon resampling curves). If our characterisation approach is
expanded to include prediction of disease (biomarker analysis), hundreds of
samples would be required to validate the predictions.
Additionally, the application of fuzzy RST would be valuable for future work.
Disease is a continuum and can rarely be described using a simple two-class
paradigm. Fuzzy RST would also remove the need to discretise the data,
preventing some information loss and resolving one of the largest drawbacks
associated with rough set characterisation.


\section*{Acknowledgment}

\noindent This work was completed under a PhD studentship supported by the 
Department for Employment and Learning (DEL) in N. Ireland. 



\bibliographystyle{IEEEtran}
\bibliography{IEEEabrv,paper}

\begin{thebibliography}{10}
\bibitem{aitchison2005compositional}
J.~Aitchison and J.~J. Egozcue, ``Compositional data analysis: where are we and
  where should we be heading?'' \emph{Mathematical Geology}, vol.~37, no.~7,
  pp. 829--850, 2005.
\end{thebibliography}

\end{document}